\documentclass[prb,twocolumn,showpacs]{revtex4}

\usepackage{graphicx}
\begin{document}

\title{
   Exact-diagonalization studies of trion energy spectra 
   in high magnetic fields}

\author{
   Arkadiusz W\'ojs,$^{1,2}$ and
   John J. Quinn$^2$}

\affiliation{
   \mbox{
   $^1$Institute of Physics, Wroclaw University of Technology,
       Wybrze\.ze Wyspia\'nskiego 27, 50-370 Wroclaw, Poland}\\
   $^2$Department of Physics, University of Tennessee, 
       Knoxville, TN 37996, USA}

\begin{abstract}
Binding energies of negative and positive trions in doped GaAs 
quantum wells in high magnetic fields are studied by exact 
numerical diagonalization in spherical geometry.
Compared to earlier calculations, finite width of the quantum 
well and its asymmetry caused by one-sided doping are both 
fully taken into account by using self-consistent subband wave 
functions in the integration of Coulomb matrix elements, and 
by inclusion of higher subbands along with several Landau levels 
in the Hilbert space.
Detailed analysis of the accuracy and convergence of the exact 
diagonalization scheme is presented, including dependence on 
Landau level and subband mixing, sensitivity to the (not well 
known) single-particle spectrum in the valence band, and the 
estimate of finite-size errors.
The main results are the exciton dispersion and trion binding 
energy spectrum calculated as a function of the magnetic field, 
quantum well width, electron concentration, and the presence 
of an ionized impurity.
As a complementary approach, a combination of the exact 
diagonalization in the quantum well plane and the variational 
calculation in the normal direction is used as well.
\end{abstract}
\maketitle

\section{Introduction}

Photoluminescence (PL) is a powerful method in the experimental 
studies of two-dimensional (2D) electron systems in high 
magnetic fields.\cite{Bassani75,Ivchenko05,Hawrylak97}
In a typical experiment, an additional electron--hole
($e$--$h$) pair is introduced into the electron system 
through absorption of a photon.
This causes energy relaxation via Coulomb scattering, 
followed by optical recombination of the well-defined,
quantized eigenstates involving an $e$--$h$ pair coupled 
in different ways to the surrounding electrons.

This coupling is a response of the 2D electron system to 
the perturbation associated with the photo-injection of a 
hole (particle of the opposite charge).
Depending on the nature of this response and on the number 
of electrons involved, the PL spectrum either reveals 
information about dynamics of an illuminated system
\cite{Heiman88} or ``merely'' probes the few-body excitonic 
complexes which form under given conditions (quantum well 
geometry, magnetic field, etc.) but are weakly affected 
by the surrounding electrons.

The latter scenario occurs at sufficiently high magnetic 
fields $B$ and low electron concentrations $\varrho$, 
corresponding to small Landau level (LL) filling factors 
$\nu=2\pi\varrho\lambda^2$ (with $\lambda=\sqrt{hc/eB}$ 
being the magnetic length).
The PL spectra at $\nu\ll1$ are usually dominated by  
recombination of three-body states called trions 
(X$^-=2e+h$), consisting of an exciton (${\rm X}=e+h$) 
bound to an additional electron.\cite{Lampert58,Kheng93}
The trion binding energy is 
\begin{equation}
   \Delta=E_{\rm X}-E_{{\rm X}^-}, 
\label{eqbnd}
\end{equation}
where $E$ is the X/X$^-$ ground state energy.
The lowest LL energy $\varepsilon_0$ is set to zero for each 
particle.

The exhaustive reviews of experimental and theoretical work on 
trions have been written by Peeters {\em et al.},\cite{Peeters01}
Bar-Joseph,\cite{BarJoseph05} and also by us.\cite{Wojs03}
Let us hence only summarize here that despite involving only 
three particles, quantum dynamics of a trion in the presence 
of confinement and magnetic field is neither trivial nor 
completely understood.
The complications include the competition of several (cyclotron, 
Coulomb, subband, and Zeeman) energy scales, complex structure
of the valence band, spin-orbit effects (and their effect on 
relaxation), and coupling to free carriers, lattice defects, 
and phonons. 

However, it is established that, depending on the parameters 
(composition and width $w$ of the quantum well, electron 
concentration $\varrho$, magnetic field $B$, etc.), the trion 
energy spectrum contains one or more bound states, distinguished
by the total spin of the pair of electrons $S$ and the relative 
angular momentum $\mathcal{M}$.
(i) At small $B$, the only bound trion is the ``singlet'' 
X$^-_{\rm s}$ with $S=0$ and $\mathcal{M}=0$, a 2D analog of 
the Hydrogen ion.\cite{Lampert58,Stebe89}
It was first observed in CdTe by Kheng {\em et al.}\cite{Kheng93} 
and then also in GaAs\cite{Buhmann95,Finkelstein95,Shields95a,%
Gekhtman96} and ZnSe\cite{Astakhov99,Homburg00} (with considerably 
different effective Rydbergs $Ry$ and Bohr radii $a_{\rm B}$).
Positive singlet trions X$^+=2h+e$ (with the opposite 
effective-mass ratio) were also observed\cite{Shields95b,Glasberg99} 
in acceptor-doped samples.
(ii) In the (unrealistic) limit of very high $B$ and vanishing $w$, 
the singlet trion unbinds as predicted from the ``hidden symmetry''
\cite{Lerner81,Dzyubenko83,Macdonald90} (particle--hole symmetry 
between conduction electrons and valence holes in the lowest LL).
However, it is replaced by a different bound state: the ``dark 
triplet'' X$^-_{\rm td}$ with $S=1$ and $\mathcal{M}=-1$,\cite{Wojs95} 
and with an infinite optical lifetime.\cite{Palacios96,Dzyubenko00} 
Despite initial difficulties\cite{Hayne99} it was also eventually 
detected in PL,\cite{Munteanu00,Yusa01,Vanhoucke01} and its vanishing 
oscillator strength was confirmed directly in optical absorption.
\cite{Schuller02} 
The singlet--triplet crossover occurs in rather high magnetic 
fields,\cite{Whittaker97} and it is fairly sensitive to the 
parameters.\cite{Wojs00,Gladysiewicz05,Redlinski01}
(iii) Additional, less strongly bound trions occur at intermediate 
fields, including the ``bright triplet'' X$^-_{\rm tb}$ with $S=1$ 
and $\mathcal{M}=0$.\cite{Wojs00}
All three trions were later confirmed by several independent 
calculations\cite{Gladysiewicz05,Riva00} and experiments.
\cite{Yusa01,Vanhoucke01,Astakhov05,Andronikov05}

Trions were also identified beyond the ``dilute regime,'' in 
PL spectra of the fractional quantum Hall states.
It was found\cite{Wojs06} that a trion immersed in a Laughlin 
$\nu=1/3$ liquid may survive in form of a fractionally 
charged ``quasiexciton'' state which retains the characteristic 
internal $e$--$h$ correlations of its trion parent.
This extends the variety of systems in which trions occur and 
also allows to use their recombination spectrum as an indirect 
probe of incompressible electron fluids.
\cite{Schuller03,Byszewski06}
This possibility seems especially attractive for systems in 
which these fluids may form but cannot be studied by transport.
\cite{Takeyama98}
Remarkably, the recombination spectra of the quasiexcitons formed 
from different trions and interacting with fractionally charged 
quasiparticles of the Laughlin liquid behave differently when the 
filling factor passes through $\nu=1/3$: they show a discontinuity 
for the X$^-_{\rm td}$ but no anomaly for the X$^-_{\rm s}$.
Therefore, the quantum wells for studying incompressible fluids 
by PL need be designed so as to yield the X$^-_{\rm td}$ ground 
state. 
This could be greatly helped by accurate theoretical predictions.

In this paper we address this problem by extensive realistic 
numerical calculations of the binding energies of (negative 
and positive) trions in doped GaAs quantum wells in high 
magnetic fields.
Compared to the earlier studies,\cite{Whittaker97,Wojs00,%
Gladysiewicz05,Redlinski01,Riva00} asymmetry of the quantum 
well caused by one-sided doping is fully taken into account 
by self-consistent calculation of the subband wave functions 
and by inclusion of up to three subbands along with several 
Landau levels in the Hilbert space.
Especially for the the singlet trion and for doped wells, 
the results are significantly different from the previous, 
less accurate models.

Another goal of this paper is to explain and justify the 
configuration-interaction exact numerical diagonalization 
in Haldane spherical geometry for the realistic calculations 
of the few-body excitonic complexes.
The method was recently used in several related studies 
without giving a detailed description,\cite{Wojs06,Bryja06} 
leaving the questions of its validity and accuracy largely 
unanswered. 
The detailed Secs.~\ref{secModel} and \ref{secCaA} fill 
this gap, and with the future studies in perspective, 
point out the advantages over the (complementary) 
variational approch.

\section{Model}
\label{secModel}

\subsection{Haldane spherical geometry}

In Haldane's spherical geometry,\cite{Haldane83} the particles 
(electrons or holes) are confined to the surface of a sphere 
of radius $R$.
The radial (i.e., normal to the surface) magnetic field $B$ is 
produced by a Dirac monopole placed at its center.
The monopole strength $2Q=4\pi R^2B/\phi_0$ is defined in the units 
of the $\phi_0=hc/e$ quantum as the total flux through the surface.
This definition can also be rewritten as $R^2=Q\lambda^2$ which makes
clear that $Q$ measures the surface curvature in the units of magnetic 
length $\lambda=\sqrt{hc/eB}$ (rather than defining a specific $B$).
Also, due to Dirac's condition, $2Q$ must be integral.

The single-particle states $Y_{Q,l,m}$ form degenerate shells of 
the eigenstates of angular momentum length $l$ and projection $m$.
They are called ``monopole harmonics.''\cite{Wu76}
The $n$th shell has $l=Q+n$ and it corresponds to the $n$th Landau 
level (LL).
Its degeneracy is $g=2l+1$ ($m$ going from $-l$ to $l$), which scales 
linearly with $B$ and $R^2$.

In the lowest LL ($n=0$), the $m=\pm Q$ states describe the closest orbits 
around the opposite poles, while, in general, $\left<z\right>\propto m$.
The wave functions scale with $\lambda$.
In the $\lambda/R\rightarrow0$ limit, the sequence of orbitals with $m=Q$, 
$Q-1$, $Q-2$, \dots, evolve continuously into the series of states with 
angular momenta $\mu=0$, $-1$, $-2$, \dots, obtained in the symmetric 
gauge on a plane.
Therefore, also the spherical orbitals with $m=Q+\mu$ can be denoted 
by the value of $\mu$ (both in the lowest the excited LLs).

The spherical geometry was originally proposed for numerical studies 
of extended systems (such as incompressible quantum liquids).
Like confinement or periodic boundary conditions, it allows for having 
a finite number of electrons (as required for the exact numerical 
diagonalization of the hamiltonian) in a finite area (to achieve a finite 
density of the actual, infinite system).
However, Haldane's geometry does it without breaking the 2D translational 
symmetry, essential for various properties of the liquid.
The 2D rotational symmetry of the sphere replaces the symmetry of the 
plane and leads to the conservation of (the same number of) two orbital 
quantum numbers -- two components of the total angular momentum: length 
$L$ and projection $L_z$.

Due to the relation between orbital Hilbert eigenspaces in both geometries, 
the ``spherical'' $L$ and $L_z$ correspond to the ``planar'' quantum numbers.
The {\em neutral} excitations, which on a plane move along straight lines 
with well defined wave vectors $k$, on a sphere move along great circles and
carry a conserved $L=kR$ (the degeneracy associated with $L_z$ corresponds
to that of the orientation of the wave vector, and the quantization of $L$
and $L_z$ is a finite size effect).
The {\em charged} excitations move along the (quantized) cyclotron orbits 
in both geometries, and $L$ must be interpreted in terms of the planar 
angular momentum $M$ rather than in terms of $k$ (in this case, finite-size 
affects the orbit by the surface curvature only).
Clearly, for the spherical geometry to be useful, the characteristic length
of correlations responsible for the studied effect must scale with 
$\lambda$, so that it can be made smaller than $R$ (and the finite size
effects can be eliminated by the $2Q\rightarrow\infty$ extrapolation).

Haldane's geometry has also become useful in studying excitons or trions.
The sole advantage is the 2D rotational symmetry of the sphere, preserving 
the degeneracy of the trion LLs in a finite size calculation.
The decoupling of trion's relative dynamics from the cyclotron motion of its 
``center of mass'' (CM) is a big help in the identification of different 
trions in the energy spectra, especially of the less strongly bound states.
We put CM in parentheses here, since it is not exactly the CM motion that 
decouples for charged complexes in a magnetic field.

Different trion states are distinguished by their relative (spin and 
orbital) correlations, reflected in difference in the relative angular 
momentum $\mathcal{M}$.
The CM excitations carry an arbitrary (negative) angular momentum 
$\mathcal{M}_{\rm cm}$ and lead to each trion having its own LL 
accommodating different total angular momenta $M=\mathcal{M}+
\mathcal{M}_{\rm cm}$.
Let us stress that the proper definition of $\mathcal{M}$ and 
$\mathcal{M}_{\rm cm}$ independently conserved in the magnetic 
field is non-trivial by involving the ``magnetic translation'' 
generators.\cite{Dzyubenko00,Avron78}

If, in order to facilitate numerical diagonalization, the single-particle 
basis is restricted to several lowest single-particle orbitals: 
$|\mu|\le\mu_{\rm max}$, then $\mathcal{M}_{\rm cm}$ and $\mathcal{M}$ 
are no longer independently conserved.
This causes artificial dependence of computed trion energies on 
$M$ and makes distinction of different trions more difficult.
This problem does not occur in spherical geometry, in which the 
trion LLs have a form of angular momentum multiplets with 
$L=Q+\mathcal{M}$.
Resolving $L$ (a conserved number regardless of finite LL degeneracy) 
in the diagonalization makes the identification simple: each bound 
$L$-multiplet is a genuine trion state.
Exact diagonalization in either geometry may be combined with the 
size extrapolation to eliminate finite-size (qualitative and 
quantitative) errors.

Let us summarize the differences of both approaches.
On the plane, the Coulomb matrix elements are accurate at each value of
$\mu_{\rm max}$, but the trion LL degeneracies are only restored in the 
$\mu_{\rm max}\rightarrow\infty$ limit.
On the sphere, the 2D symmetry holds at each $2Q$, but the interaction 
matrix elements depend on surface curvature.
The last effect can be reduced by calculation of matrix elements 
from Haldane pseudopotentials for a plane (using the spherical 
expansion and Clebsch-Gordan coefficients), helpful when the 
$2Q\rightarrow\infty$ extrapolation is impossible.

\subsection{Hamiltonian matrix elements}

\subsubsection{Basis}

For diagonalization of the hamiltonian $H$ we choose the configuration 
interaction (CI) basis, e.g., 
\begin{equation}
   \left|i,j;k\right>=
   c_i^\dagger c_j^\dagger d_k^\dagger \left|{\rm vac}\right>
\end{equation}
for the trion.
Here $c^\dagger$ and $d^\dagger$ are electron and hole creation operators, 
and the composite indices $i$, $j$, and $k$ include all relevant 
single-particle quantum numbers (spin $\sigma$, well subband $s$, 
LL index $n$, and angular momentum $m$).
Although for only two or three particles the CI basis does not produce as 
sparse matrices of the (two-body) hamiltonian as for many-electron systems,
it is still a convenient choice for the evaluation of matrix elements.
The single-particle states used in the exciton and trion basis include 
both spin projections ($\sigma=\uparrow$ and $\downarrow$), up to five 
lowest LLs ($n\le4$) and three lowest subbands ($s\le2$) for electrons 
and holes.
Different values of $2Q\le30$ were used to eliminate dependence on finite 
LL degeneracy.

The pair electron spin $S$ (for the trion) and total angular momentum 
$L$ as well as their projections $S_z=\sum\sigma$ and $L_z=\sum m$ are 
all conserved quantum numbers.
The CI basis is the eigenbasis of $S_z$ and $L_z$, while the energy 
(excluding the Zeeman term) only depends on $S$ and $L$.
Therefore, only the subspace corresponding to the minimum values of $S_z=0$
and $L_z=0$ (or $1/2$, if half-integral) need be diagonalized, 
containing all $S$ and $L$ states and thus covering the entire energy 
spectrum.

\subsubsection{One-body terms}
\label{sec1BT}

The exact cyclotron energy on the sphere is 
\begin{equation}
   \varepsilon_n=
   \hbar\omega_c\left(n+{1\over2}+{n(n+1)\over2Q}\right).
\end{equation}
In calculation, the last term equal to $n(n+1)\hbar^2/2J$
(here, $J=m^*R^2$ is the moment of inertia on a sphere, 
and $m^*$ is the effective mass, not to be confused with 
angular momentum) was omitted as a finite-size artifact, 
and the realistic values of $\hbar\omega_c$ for GaAs were 
taken from experiment.
For electrons, $\hbar\omega_c/B=1.78$~meV/T; for the heavy 
holes we used the following interpolated formula 
\begin{equation}
   \hbar\omega_c=
   \alpha\,(1+\beta_1w^{-2})+\gamma\,(1+\beta_2w^{-2})B, 
\label{eq0}
\end{equation}
with the following parameter set: $\alpha=0.45$~meV, 
$\gamma=0.282$~meV/T, $\beta_1=275$~meV\,nm$^2$ and 
$\beta_2=10$~meV\,nm$^2$, taken after Cole {\em et al.}
\cite{Cole97} and expected to work adequately at $B\ge10$~T 
in $w=10-30$~nm wells (in doped wells, the actual effective 
width of the hole layer is used for $w$).

Although Eq.~(\ref{eq0}) describes well only the $n=0\rightarrow1$ gap, 
we used it (for the lack of better information) to model scattering 
to higher LLs as well, and tested sensitivity to this approximation.
We also neglected the mixing of heavy and light holes (off-diagonal 
terms of the Luttinger-Kohn hamiltonian).
We expect these approximations to affect the negative trion binding 
energy $\Delta$ much less than each of the exciton and trion energies, 
$E_{\rm X}$ and $E_{{\rm X}^-}$, separately.
However, a more accurate treatment of the hole LL spectrum might 
considerably improve the model for the positive trions. 

The energies of higher quantum well subbands (excited states in the 
$z$-direction) $\mathcal{E}_s$ were calculated as a function of well 
width $w$ and electron concentration in the well $\varrho$.
In doped wells ($\varrho>0$), the electric field of the remote 
ionized donors penetrates the quantum well, affecting the 
positions of occupied energy levels in the well (for our parameters, 
it is only $\mathcal{E}_0$) with respect to the Fermi energy.
This in turn affects $\varrho$ and the sheet concentration of 
ionized donors, and thus also the electric field itself.
The Schr\"odinger and Poisson equations must be therefore solved 
self-consistently to give the values of $\mathcal{E}_s$.\cite{Tan90}

As an example, let us look at the values for the $s=1$ subband 
(we set $\mathcal{E}_0=0$), which for $w=10-30$~nm can be fitted 
by power-law curves $\mathcal{E}_1=\zeta\,w^{-\kappa}$.
For the electrons we obtain $\zeta=3600$ and $\kappa=1.60$, weakly dependent 
on doping, and for the holes, $\zeta=1400$ and $\kappa=1.87$ for a symmetric 
well and $\zeta=430$ and $\kappa=1.25$ for $\varrho=2\cdot10^{11}$~cm$^{-2}$ 
(the values appropriate for $\mathcal{E}$ measured in meV and and for $w$ 
measured in nm).
Clearly, strong subband mixing can be anticipated at least in 
higher fields and in wider wells (e.g., compare $\mathcal{E}_1=30.5$ and 
6.9~meV to $\hbar\omega_c=44.5$ and 7.6~meV for electron and hole, 
respectively, in a symmetric 20~nm well at $B=25$~T). 

Since the projection of pair electron spin, $S_z$, is conserved by the 
Coulomb interaction, the Zeeman energy $E_{\rm Z}$ does not have to be 
included in the diagonalization.
Instead, the entire energy spectra obtained from the diagonalization 
without $E_{\rm Z}$ must be at the end rigidly shifted by $S_z E_{\rm Z}$.
For the trion binding energies it means subtracting one electronic 
$E_{\rm Z}$ from $\Delta$ of the singlet states (as needed to flip the 
spin of one of its two unpolarized electrons).
In this approximation, the spin-orbit coupling is neglected and it is
assumed that that electron and hole $E_{\rm Z}$'s are both constants.
In fact, they depend on $n$ and $s$, through Rashba and (in asymmetric 
structures) Dresselhaus couplings, both mixed by Coulomb 
interaction.\cite{Zawadzki04} 

In the recombination spectrum of excitons or trions, the sum of electron
and hole Zeeman energies splits the ``$\pm1$'' peaks corresponding to
two circular polarizations of light (both present if both electron and 
hole spins are populated).
If $E_{\rm Z}$ were constant, the same splitting would result for the 
X and all X$^-$ states.
The observed difference\cite{Glasberg99,Gladysiewicz06} reveals the 
inaccuracy of this approximation.
It also reflects different LL and subband mixing in each of these 
complexes.
In our model, the spin-orbit effects can be included as the zeroth 
order perturbation, given $n$- and $s$-dependence of $E_{\rm Z}$ 
(the LL and subband occupation numbers are easily extracted from 
the X and X$^-$ wave functions obtained from the diagonalization).

\subsubsection{Two-body terms}

\paragraph{Ideal 2D layers.}

Let us begin with derivation of the matrix elements of the 
Coulomb interaction $V_{ee}\equiv V$,
\begin{equation}
   V(r)={e^2\over\epsilon r}
\label{eq1}
\end{equation}
(with $\epsilon$ being the dielectric constant) for the layers 
with zero width.
We use rays in spherical coordinates $\Omega=(\theta,\phi)$
as the arguments of monopole harmonics and write
\begin{eqnarray}
   &&
   \left<\sigma_1',n_1',m_1';\sigma_2',n_2',m_2'|V|
         \sigma_1 ,n_1 ,m_1 ;\sigma_2 ,n_2 ,m_2 \right>
\nonumber\\
   &=&
   \delta_{\sigma_1}^{\sigma_1'}
   \delta_{\sigma_2}^{\sigma_2'}
   \int\! d^2\Omega_1\, d^2\Omega_2\,
   Y_{Q,l_1',m_1'}^*(\Omega_1)\,
   Y_{Q,l_2',m_2'}^*(\Omega_2)
\nonumber\\
   &&\rule{18mm}{0mm}
   \times V(r)\,
   Y_{Q,l_1,m_1}(\Omega_1)\,
   Y_{Q,l_2,m_2}(\Omega_2).
\end{eqnarray} 
This four-dimensional integral is solved by using the 
rotation relations for monopole harmonics,\cite{Wu77}
\begin{equation}
   \hat\mathcal{D} \, Y_{Q,l,m}(\Omega)
   = e^{i\psi} \sum_\eta \mathcal{D}^{\,l}_{\eta,m} \, 
   Y_{Q,l,\eta}(\Omega),
\label{eq2}
\end{equation}
different from those for standard spherical harmonics only 
by an inconsequential phase factor $\psi(Q)$.
The Euler rotation\cite{Sakurai94} 
$\hat\mathcal{D}(0,-\theta_1,-\phi_1)$ moves $\Omega_1$ 
to the north pole of the sphere $\hat N$.
Its Wigner function is
\begin{equation}
   \mathcal{D}_{\eta,m}^{\,l} \equiv
   \mathcal{D}_{\eta,m}^{\,l}(0,-\theta_1,-\phi_1) =
   e^{im\phi_1} d_{\eta,m}^{\,l}(-\theta_1),
\label{eq3}
\end{equation}
where we use standard notation for the reduced function
\begin{equation}
   d_{\eta,m}^{\,l}(-\theta_1)=
   (-1)^p\sum_\tau \xi_{\eta,m}^{\,l,\tau} \, 
   u_1^{2l} \, (v_1/u_1)^{2\tau+p}
\label{eq4}
\end{equation}
with $u_1=\cos(\theta_1/2)$, $v_1=\sin(\theta_1/2)$, $p=\eta-m$, 
\begin{equation}
   \xi_{\eta,m}^{\,l,\tau} =
   (-1)^{\tau+p}
   \sqrt{{l+\eta\choose\tau+p}{l-\eta\choose\tau }
         {l+m   \choose\tau  }{l-m   \choose\tau+p}},
\label{eq5}
\end{equation}
and the summation going over all $\tau$ for which none 
of the arguments of the factorials in $\xi$ are negative.

In Eq.~(\ref{eq1}), we take the chord distance 
$r=(\Omega_1-\Omega_2)R$.
From now on, let us use $\theta\equiv\theta_{12}$ for the 
angle between $\Omega_1$ and $\Omega_2$.
The Coulomb potential becomes
\begin{equation}
   V(\theta)={e^2\over2\epsilon Rv}.
\label{eq5a}
\end{equation}
By applying rotation $\mathcal{D}$ under the integral over 
$\Omega_1$ we get
\begin{eqnarray}
   &&\left<n_1',m_1';n_2',m_2'|v^{-1}|
         n_1 ,m_1 ;n_2 ,m_2 \right>
\label{eq6}
\\
   &&=
   \sum_\eta 
   \left<n_1',m_1'\right|
   \mathcal{D}_{\eta,m_2'}^{\,l_2'*} 
   \mathcal{D}_{\eta,m_2 }^{\,l_2  } 
   \left|n_1,m_1\right>
   \left<n_2',\eta\right|
   v^{-1}
   \left|n_2,\eta\right>.
\nonumber
\end{eqnarray}
Using Eqs.~(\ref{eq3}--\ref{eq5}), the first matrix 
element above is
\begin{eqnarray}
   &&\left<n_1',m_1'\right|
   \mathcal{D}_{\eta,m_2'}^{\,l_2'*} 
   \mathcal{D}_{\eta,m_2 }^{\,l_2  } 
   \left|n_1 ,m_1 \right>
\nonumber\\
   &&=
   \delta^{m_1'+m_2'}_{m_1+m_2}
   \sum_{\alpha',\alpha}
   \xi_{\eta,m_2'}^{\,l_2',\alpha'} 
   \xi_{\eta,m_2 }^{\,l_2 ,\alpha } 
   I^{n_1',m_1'}_{n_1,m_1}(a,b)
\label{eq7}
\end{eqnarray}
with $b=2(\alpha'+\alpha+\eta)-m_2'-m_2$, $a=2(l_2'+l_2)-b$,
and
\begin{equation}
   I^{n',m'}_{n,m}\!(a,b)= 2\pi\!\! \int\!\! d\cos\theta\,\,
   Y_{Q,l',m'}(\theta)\,u^av^b\,Y_{Q,l,m}(\theta).
\label{eq8}
\end{equation}
The second integral in Eq.~(\ref{eq6}) is, up to a prefactor,
an effective one-body Coulomb pseudopotential 
\begin{equation}
   \mathcal{V}_n^{n'}(m)
   \equiv\left<n',m\right|V(\theta)\left|n,m\right>
\end{equation}
describing scattering of an electron or a hole by a point 
charge (e.g., ionized impurity) located at the north pole.
Note that it is different from the definition of Haldane 
(pair) pseudopotential.
The $\mathcal{V}(m)$ for the sphere can be converted into 
$\mathcal{V}(\mu)$ for the plane via $\mu=m-Q$.
This interaction conserves $m$, but allows for an inter-LL 
transition $n\rightarrow n'$.
Using the symbol $I$ it simply reads
\begin{equation}
   \left<n_2',\eta\right|v^{-1}\left|n_2,\eta\right>
   =I^{n_2',\eta}_{n_2,\eta}(0,-1).
\label{eq8a}
\end{equation}
To calculate the integral $I$ of Eq.~(\ref{eq8}) 
let us recall that 
\begin{equation}
   Y_{Q,l,m}(\theta)= 
   N_{Q,l,m} u^{Q+m} v^{Q-m} P_n^{Q-m,Q+m}(\cos\theta),
\label{eq9}
\end{equation}
where 
\begin{equation}
   N_{Q,l,m}^2 ={2l+1\over4\pi}{2l\choose l+m}{2l\choose l+Q}^{-1},
\label{eq10}
\end{equation} 
and the Jacobi polynomial $P$ can be expanded as
\begin{equation}
   P^{\alpha,\beta}_n(x)=
   \sum_q c^{\alpha,\beta}_{n,q} \, u^{2(n-q)} \, v^{2q},
\label{eq11}
\end{equation}
with the coefficients
\begin{equation}
   c^{\alpha,\beta}_{n,q}
   =(-1)^q {n+\alpha\choose n-q} {n+\beta\choose q},
\label{eq12}
\end{equation}
and the summation range defined by the above factorials.

Finally, substitution of Eqs.~(\ref{eq9}--\ref{eq12}) 
into Eq.~(\ref{eq8}) gives
\begin{eqnarray}
   I^{n_1,m_1}_{n_2,m_2}(a,b)
   &=& 
   N_{Q,l_1,m_1}
   N_{Q,l_2,m_2}
\label{eq13}
\\
   &\times&
   \sum_{q_1,q_2} c^{Q-m_1,Q+m_1}_{n_1,q_1}
                  c^{Q-m_2,Q+m_2}_{n_2,q_2} 
   N_{\tilde{Q},Q,\tilde{m}}^{-2},
\nonumber
\end{eqnarray}
where $\tilde{Q}=Q+a+b+(n_1+n_2)/2$ and $\tilde{m}=a-b+
(m_1+m_2+n_1+n_2-q_1-q_2)/2$.

By collecting all above equations, a closed expression
for the $e$--$e$ matrix element in a layer of zero width 
is finally obtained.
The $e$--$h$ elements are found by particle-hole 
conjugation (with $i$ below being composite indices)
\begin{equation}
   \left<i_1';i_2'|V_{eh}|i_1;i_2 \right>=
   -\left<i_1';i_2 |V_{ee}|i_1;i_2'\right>.
\label{eq14}
\end{equation}

\paragraph{Layers of finite width.}
\label{secLFW}

In the ideal 2D limit, the interaction matrix $V$ depended on $B$ 
only through the overall scale $e^2/\lambda$.
A finite width $w$ of the quantum well, causing finite extent 
of wave functions in the normal ($z$) direction, introduces an 
additional parameter $w/\lambda\propto\sqrt{B}$ that affects 
matrix elements individually.
This is true even in the simplest case when the quantum well 
asymmetry and the difference between electron and hole subband 
wave functions can be neglected.
Hence, the matrix elements no longer explicitly depend only 
on $2Q$ (surface curvature), but also on the form of subband 
wave functions and on $B$ (via the $w/\lambda$ length scale 
ratio).

\begin{figure}
   \includegraphics[width=3.4in]{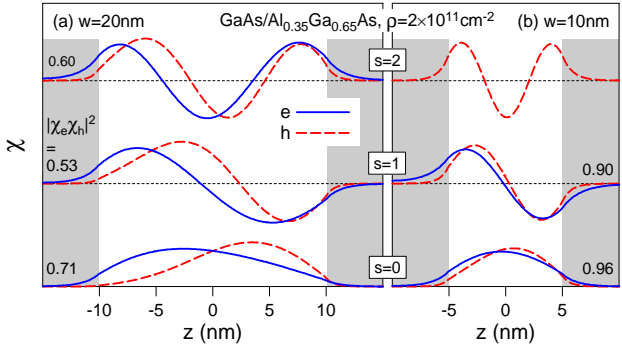}
   \caption{(color online)
      Lowest electron and hole subband wave functions 
      $\chi_s(z)$ in GaAs quantum wells of width $w=10$~nm 
      (a) and 20~nm (b) doped on one side to electron 
      concentration $\varrho=2\cdot10^{11}$~cm$^{-2}$.
      Squared electron-hole overlaps are indicated.
      Note only two bound electron subbands in (b).}
   \label{fig01}
\end{figure}

The realistic electron and hole subband wave functions $\chi_s(z)$ 
for asymmetrically doped quantum wells were calculated 
self-consistently\cite{Tan90} (together with the subband energies 
$\mathcal{E}_s$ of Sec.~\ref{sec1BT}).
An example is shown in Fig.~\ref{fig01} for two GaAs quantum wells 
of different widths, both doped on one side to the same electron 
concentration $\varrho=2\cdot10^{11}$~cm$^{-2}$.
The charge separation due to an electric field inside the well
is significant for the $w=20$~nm well and negligible for $w=10$~nm.
This is also confirmed by the indicated squared $e$--$h$ overlaps
$\left|\left<\chi_e|\chi_h\right>\right|^2$.

To include the finite quantum well width in Coulomb matrix elements, 
the bare potential $V(r)$ must be replaced by an effective in-plane 
potential containing the appropriate form factors,
\begin{equation}
   V^{s_1',s_2'}_{s_1,s_2}(r)= 
   {e^2\over\epsilon} \int dz_1 dz_2
  {\chi_{s_1'}^*(z_1) \, \chi_{s_2'}^*(z_2)\,
   \chi_{s_1 }  (z_1) \, \chi_{s_2 }  (z_2)
   \over\sqrt{r^2+(z_1-z_2)^2}}.
\label{eq15}
\end{equation}
On a sphere, $V^{s_1',s_2'}_{s_1,s_2}(\theta)$ is obtained from the 
above by replacing $r$ with $2Rv$, exactly as in Eq.~(\ref{eq5a}) 
for the ideal 2D layers.
In the matrix element $\left<i_1';i_2'|V|i_1;i_2\right>$ the first 
integral in Eq.~(\ref{eq6}) in unchanged, and the modified potential 
$V(\theta)$ enters only the one-body pseudopotential $\mathcal{V}$, 
which is hence no longer given by Eq.~(\ref{eq8a}).
However, since $\chi_e\ne\chi_h$, different effective potentials 
$V$ and pseudopotentials $\mathcal{V}$ must be used for $e$--$e$ 
and $e$--$h$ interactions.

Not being able to calculate $\mathcal{V}$ analytically, we are 
forced to compute the entire tables of these 3D integrals for 
each choice of subband wave functions $\chi$ and $2Q$,
\begin{equation}
   \mathcal{V}_{s_1,s_2,n}^{s_1',s_2',n'}(m)
   =\left<n',m\right|
   V^{s_1',s_2'}_{s_1,s_2}(\theta)
   \left|n,m\right>.
\label{eq16}
\end{equation}
The above pseudopotential describes inter-LL scattering of an 
electron or a hole by an artificial charge that is localized 
at the north pole within the 2D sphere and undergoes transition 
between the subband states $s_1$ and $s_1'$ along the normal 
direction.
It also can be converted into $\mathcal{V}(\mu)$ for the plane
via $\mu=m-Q$.

\begin{figure}
   \includegraphics[width=3.4in]{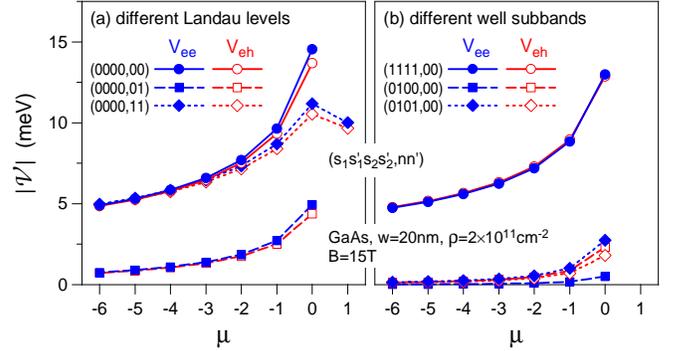}
   \caption{(color online)
      Examples of one-body pseudopotentials (interaction energy 
      $\mathcal{V}$ vs angular momentum $\mu$) defined by 
      Eq.~(\ref{eq16}) and used in calculation of two-body
      matrix elements for $e$--$e$ and $e$--$h$ Coulomb 
      scattering in/between different LLs $n$ (a) and subbands 
      $s$ (b), calculated for one-sided doped 20~nm GaAs 
      quantum well at magnetic field $B=15$~T.}
   \label{fig02}
\end{figure}

Several examples of $\mathcal{V}(\mu)$ are plotted in 
Fig.~\ref{fig02} to illustrate the dependence on subband 
and LL indices.
Despite a significant charge separation apparent in 
Fig.~\ref{fig01}(a), the corresponding $e$--$e$ and $e$--$h$ 
pseudopotentials are rather similar.
However, the difference reaching almost 1~meV (for $w=20$~nm, 
$\varrho=2\cdot10^{11}$~cm$^{-2}$, and $B=15$~T) is sufficient 
to have a big effect on the trion binding energy.

Scattering by a point charge (ionized impurity) at a 
distance $d$ from the middle of the well is described by 
a different pseudopotential,
\begin{equation}
   \mathcal{V}_{s,n;\,d}^{s',n'}(m)
   =\left<n',m\right|V^{s'}_{s;\,d}(\theta)\left|n,m\right>,
\label{eq17}
\end{equation}
with the following effective potential,
\begin{equation}
   V^{s'}_{s;\,d}(r)= {e^2\over\epsilon} \int dz
   {\chi_{s'}^*(z)\chi_s(z)\over\sqrt{r^2+(z-d)^2}},
\end{equation}
where, as before, $r$ is replaced by $2Rv$ for the sphere,
and conversion to $\mathcal{V}(\mu)$ is immediate.

\begin{figure}
   \includegraphics[width=3.4in]{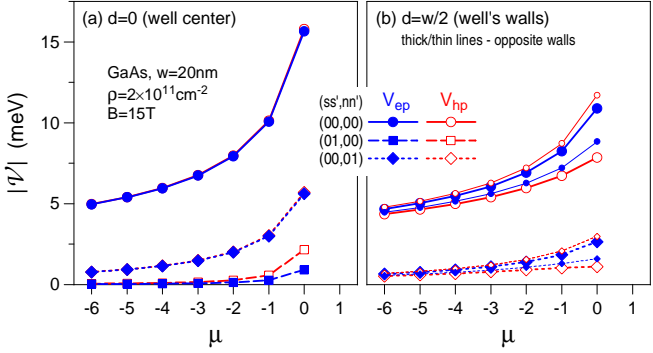}
   \caption{(color online)
      Pseudopotentials similar to Fig.~\ref{fig02}, but 
      defined by Eq.~(\ref{eq17}) and describing interaction 
      of an electron $e^-$ or a hole $h^+$ with a point charge 
      $p^\pm$ located in the center (a) or at either wall (b) 
      of the quantum well.}
   \label{fig03}
\end{figure}

The examples in Fig.~\ref{fig03} illustrate the sensitivity
to the impurity position $d$ and to the electron or hole 
orbital indices $s$ and $n$, as well as the electron/hole
asymmetry.
Remarkably, the energy scale of carrier--impurity interaction 
is not much (though nevertheless by up to 2~meV) greater than 
of the carrier--carrier interaction of Fig.~\ref{fig02}.

The computation of a great number of needed integrals 
(\ref{eq16}) can be accelerated by an adiabatic approximation 
in which the 3D integration is divided into the averaging of 
a squared relative $z$-coordinate 
\begin{equation}
   (\bar{D}^{s_1',s_2'}_{s_1,s_2})^2=
   \left<s_1',s_2'|(z_1-z_2)^2|s_1,s_2\right>,
\label{eqavd}
\end{equation}
and integration (only over $\theta$) of an in-plane potential
\begin{equation}
   \bar{V}^{s_1',s_2'}_{s_1,s_2}(r)={e^2/\epsilon \over
   \sqrt{r^2+(\bar{D}^{s_1',s_2'}_{s_1,s_2})^2}}.
\end{equation}
As its previous versions,\cite{Wojs00,Chapman97} the 
adiabatic approximation relies on $\bar{D}^2$ being small 
compared to $\left<n',m|r^2|n,m\right>$, so its accuracy 
depends on the in-plane quantum numbers.
We found that, for a given combination of $s$'s and $n$'s, 
it is satisfactory for sufficiently small $m$. 
However, the exact 3D integration must still be carried out 
for those several largest $m$'s (e.g., we chose $Q-m=-\mu
\le5$).

\begin{figure}
   \includegraphics[width=3.4in]{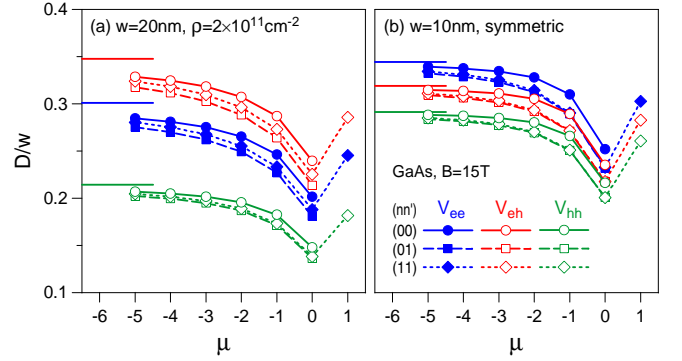}
   \caption{(color online)
      Effective $e$--$e$, $e$--$h$, and $h$--$h$ separations
      $D$ in direction $z$ normal to the quantum well plane
      defined by Eq.~(\ref{eq18}) as a function of relative
      angular momentum $\mu$.
      Horizontal lines mark averages $\bar{D}$ of Eq.~(\ref{eqavd}).
      Different LL combinations $(nn')$ are shown, 
      for the lowest subband only.
      (a) Same system as in Fig.~\ref{fig02};
      (b) symmetric or undoped 10~nm quantum well.
      Magnetic field $B=15$~T in both frames.}
   \label{fig04}
\end{figure}

A correct effective average separation $D$ in the normal 
direction (such as to give the exact value of $\mathcal{V}$) 
can be defined for each set of $s$'s, $n$'s, and $m$ by the 
requirement that (for this particular choice of quantum numbers)
\begin{equation}
   \mathcal{V}_{s_1,s_2,n}^{s_1',s_2',n'}(m)=
   \left<n',m\right|
   {e^2/\epsilon\over
   \sqrt{r^2+(D^{s_1',s_2';\,n'}_{s_1,s_2;\,n,m})^2}}
   \left|n,m\right>.
\label{eq18}
\end{equation}
Certainly, $D^{s_1',s_2';\,n'}_{s_1,s_2;\,n,m}\approx 
\bar{D}^{s_1',s_2'}_{s_1,s_2}$ for sufficiently small $m$,
while it begins to depend on $n'$, $n$, and $m$ as $m$
increases and approaches $Q$ (i.e., as $\mu=m-Q$ approaches 
zero).
The typical dependences are shown in Fig.~\ref{fig04}, with 
the averages of Eq.~(\ref{eqavd}) also marked by horizontal 
lines.
Evidently, $D$ depends on $n$ (or $s$, not shown) much less
than $\mathcal{V}$ does.
On the other hand, using a constant value $\bar{D}$ for all
values of $\mu$ is not a justified approximation.

When the whole table of 3D integrals $\mathcal{V}$ is known, 
the corresponding table of $D$'s can be immediately calculated 
by numerical solution of the 1D equations (\ref{eq18}).
We found that the values of $D(\mu)$ do not depend on the 
surface curvature nearly as much as $\mathcal{V}(\mu)$.
Therefore, for a given quantum well (the width $w$ and the 
set of subband wave functions $\chi$) and magnetic field $B$, 
the time-consuming exact 3D integration of the $\mathcal{V}$ 
table need only be performed for one value of $2Q$ and for 
several largest values of $\mu$ (e.g., we used $2Q=20$ and
$\mu\ge-5$).
The approximate (but accurate) tables of $\mathcal{V}$ for 
other values of $2Q$ are then quickly recovered from the 
corresponding table $D$, quickly computed from either 
Eq.~(\ref{eqavd}) or (\ref{eq18}), depending on $\mu$.
Finally, the two-body Coulomb matrix is calculated (analytically) 
from the coefficients $\mathcal{V}$ in the same way as 
shown in Sec.~\ref{secLFW} for the ideal 2D layer.

\subsection{Exact diagonalization}

The hamiltonian matrices $H$ were diagonalized numerically
using a Lanczos based algorithm.
The matrix was stored in a packed ``compressed-row'' format
and divided into a number of segments that could fit in 
computer's memory.
The segments were loaded one by one during the matrix-vector 
multiplication (the only operation performed on the matrix
-- once in each Lanczos iteration).
Since the vanishing of a matrix element $\left<i',j';k'|H
|i,j;k\right>$ depends only on the two basis states, the 
location of non-zero elements (the most time-consuming part) 
had to be only done once
for all used combinations of $w$, $\varrho$, or $B$.

To resolve all quantum numbers conserved by $H$ and obtain 
the energy spectrum in the form of $E(S_z,S,L_z,L)$, the 
Lanczos procedure must be carried out inside the eigensubspaces
of pair electron spin $S$ (for the trion) and total angular 
momentum $L$ (note that $S_z$ and $L_z$ are automatically 
resolved in the CI basis).
The matrix elements of ladder operators $S^+=\sum\sigma^+$ 
and $L^+=\sum l^+$ (summations go over all particles) have 
a simple form in the CI basis, allowing analytic expression
of the $\hat{L}^2=L^+L^-+L_z(L_z-1)$ and $\hat{S}^2=S^+S^-
+S_z(S_z-1)$ matrices.

The initial Lanczos $LS$-eigenstates with desired 
eigenvalues of $\hat{L}^2$ and $\hat{S}^2$ were 
calculated using a conjugate-gradient method.\cite{Paige75}
To avoid dealing with two independent eigenproblems, 
we used a joint operator $\hat{S}^2+2\hat{L}^2$, whose 
eigenstates are simultaneous and unique eigenstates 
of both $L$ and $S$ (owing to their discrete spectra).

The accumulation of numerical errors leading to escape
from the initially set $LS$ subspace after more than a few
Lanczos iterations was remedied by the $LS$-projection of 
each consecutive Lanczos state.
When more than one lowest eigenstate was needed at some
$L$ and $S$, orthogonalization to all preceding Lanczos 
states was done at every iteration to avoid ``ghost states.''
Only three most recent Lanczos states were held in computer's 
memory at the same time; all earlier ones were stored on a 
disk, in need of either the above-mentioned orthogonalization 
or recalculation of the eigenstates into the original CI basis.

Depending on spin and orbital symmetry, the diagonalization 
need not be repeated in every $(S_z,S,L_z,L)$ subspace.
In the absence of an impurity, it is only done for $L_z=0$ 
(or $1/2$ if half-integral).
Furthermore, the spin quantum numbers are only relevant for the 
trions, and their calculation was only carried out for $S_z=0$.

\section{Convergence and accuracy}
\label{secCaA}

Being essentially a perturbative method, the CI exact 
diagonalization converges as a function of the size of the 
many-body CI matrix.
Fast convergence requires that interactions are weak compared 
to the single-particle gaps.
On the other hand, the maximum size of the CI matrix is limited 
by both the knowledge of low-energy single-particle spectra and 
by the computational capability.

\begin{table}
\caption{
   Dimensions of the $2e+h$ configuration-interaction bases and 
   corresponding numbers of above-diagonal non-zero matrix 
   elements of the hamiltonian $H$, squared angular momentum 
   $\hat{L}^2$, and squared $2e$ spin $\hat{S}^2$, for the 
   $S_z=L_z=0$ subspace and for different monopole strengths 
   $2Q$ and maximum subband and LL indices $s_{\rm max}$ and 
   $n_{\rm max}$.}
\begin{ruledtabular}
\begin{tabular}{rcc|rrrr}
   $2Q$ & $s_{\rm max}$ & $n_{\rm max}$ &
   dimension & $N_H$ & $N_L$ & $N_S$ \\
\hline
   20 & 0 & 0 & 331 & 7620 & 
                2191 & 640 \\
   20 & 0 & 4 & $0.6\cdot10^5$ & $0.4\cdot10^8$ &
                $0.4\cdot10^6$ & $1.1\cdot10^5$ \\
   30 & 0 & 4 & $1.1\cdot10^5$ & $1.1\cdot10^8$ &
                $0.8\cdot10^6$ & $2.3\cdot10^5$ \\
   20 & 1 & 4 & $4.6\cdot10^5$ &$13.5\cdot10^8$ &
                $3.1\cdot10^6$ & $9.2\cdot10^5$ \\ 
   20 & 2 & 2 & $2.9\cdot10^5$ & $6.5\cdot10^8$ &
                $1.9\cdot10^6$ & $5.7\cdot10^5$
\end{tabular}
\end{ruledtabular}
\label{tab1}
\end{table}

In Tab.~\ref{tab1} we list sizes of the $H$, $L^2$, and 
$S^2$ matrices for several CI bases used in computation.
The $2e+h$ hamiltonians are not very sparse 
($2N_H/{\rm dim}>1\%$),
so numerical feasibility is defined almost exclusively by 
the number of their non-zero elements $N_H$ (rather than 
by the space dimension or the sparsity of $L^2$ or $S^2$).
For the largest basis we have used, compressed storage of 
a single $H$-matrix required over 15~GB of disk space.

Constrained by the manageable-size restriction and by limited 
knowledge of the matrix elements, we are still able to calculate 
some quantities with useful accuracy and to model various 
dependences at least qualitatively.
Below we present several tests carried out to estimate convergence 
of our calculations of the trion binding energies $\Delta$.

\begin{figure}
   \includegraphics[width=3.4in]{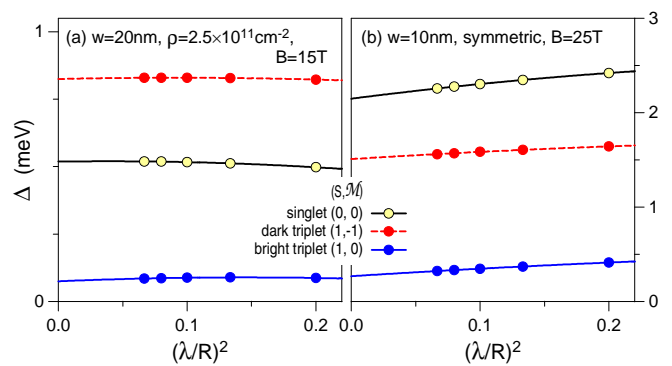}
   \caption{(color online)
      Dependence of binding energies $\Delta$ of different 
      trions X$^-$ (distinguished by the $2e$ spin $S$ and relative 
      angular momentum $\mathcal{M}$) on squared surface 
      curvature ($R$ is the sphere radius; $\lambda$ is 
      the magnetic length), calculated including five LLs and the 
      lowest subband only for (a) one-sided doped 20~nm GaAs 
      quantum well at magnetic field $B=15$~T, and (b) 
      symmetric/undoped 10~nm well at $B=25$~T.}
   \label{fig05}
\end{figure}

First, we analyze the finite-size and surface-curvature errors.
In Fig.~\ref{fig05} we plot $\Delta$ calculated in two different 
systems (a narrow symmetric well in a high magnetic field and
a wider asymmetric well in a lower magnetic field), including 
one subband and five LLs, as a function of the surface curvature 
$(\lambda/R)^2=Q^{-1}$.
The dependence is fairly weak, justifying the use of spherical 
geometry for the quantitative calculation.
Moreover, the regular behavior of $\Delta(Q^{-1})$ allows 
reliable extrapolation to the (planar) $R\rightarrow0$ limit.
In the following we used $2Q=20$.

\begin{figure}
   \includegraphics[width=3.4in]{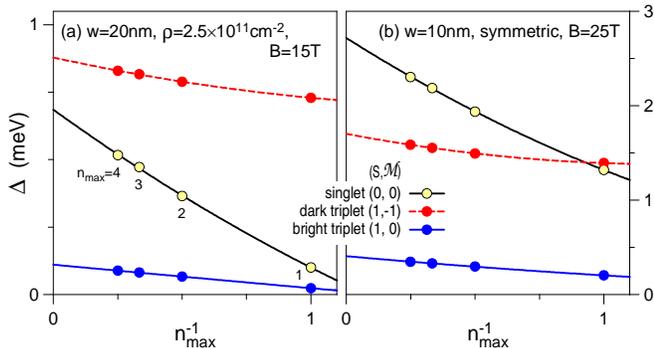}
   \caption{(color online)
      Similar to Fig.~\ref{fig05} but with dependence on
      inverse maximum LL index $n_{\rm max}$ (number of 
      included LLs) calculated for $2Q=20$ and including 
      the lowest subband only.}
   \label{fig06}
\end{figure}

In Fig.~\ref{fig06} we study dependence on the number of 
included LLs.
Although electron and hole cyclotron energies are different, 
we used the same maximum value $n_{\rm max}$ for both particles, 
because the rotational symmetry of a sphere (or the magnetic 
translational symmetry of a plane) forbids mixing of different 
electron and hole LLs in the optically active ($k=0$) exciton 
ground state: $\left<n_e',m_e';n_h',m_h'|V|0,m_e;0,m_h\right>
\propto\delta_{n_e'n_h'}$.
The dependence of $\Delta$ on $n_{\rm max}$ is much stronger 
for the singlet trion than for both triplets.
This is due to the stronger Coulomb interaction in this most 
compact state (the spin-unpolarized electrons have a large 
projection on the pair state with the minimum relative angular 
momentum $\mu=0$, not allowed for in a triplet state), causing
a more efficient mixing of different LLs.
In the following we included five LLs ($n_{\rm max}=4$).
This promises to describe fairly accurately both triplet states,
but $\Delta$ for the singlet state will probably be noticeably 
underestimated (by ~0.2 and 0.4~meV in the two shown examples).

\begin{figure}
   \includegraphics[width=3.4in]{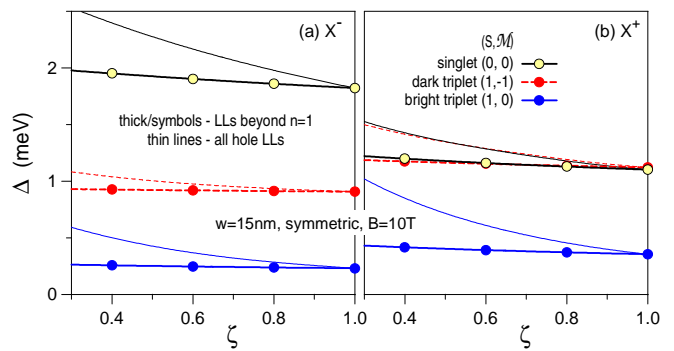}
   \caption{(color online)
      Dependence of negative (a) and positive (b) trion binding 
      energies $\Delta$ on factor $\zeta$ reducing the hole cyclotron
      gaps compared to Eq.~(\ref{eq0}), calculated for 
      symmetric/undoped 15~nm quantum well at magnetic field $B=10$~T, 
      including five LL and two subbands.
      Thin lines:
      $\varepsilon_n=\varepsilon_0+$ $n\,\zeta\hbar\omega_c$ ($n\ge1$);
      thick lines and symbols:
      $\varepsilon_1=\varepsilon_0+\hbar\omega_c$ and
      $\varepsilon_n=\varepsilon_1+(n-1)\,\zeta\hbar\omega_c$ ($n\ge2$),
      i.e., only higher gaps reduced.}
   \label{fig07}
\end{figure}

Inclusion of a great number of LLs is more difficult for the 
holes due to the complicated structure of the valence band.
Since the correct energies $\varepsilon_n$ of higher LLs are 
not known in general (for arbitrary $w$, $\varrho$, and $B$), 
we largely ignore this complication by assuming equidistant 
LLs separated by the cyclotron energy taken from experiment 
for the lowest ($n=0\rightarrow1$) transition.
In Fig.~\ref{fig07} we test sensitivity of the results to
this assumption by studying the dependence of $\Delta$ on
the dimensionless factor $\zeta$ reducing the select ($n$th) 
hole's cyclotron gaps $\Delta\varepsilon_n=\varepsilon_{n+1}
-\varepsilon_n$ compared to $\hbar\omega_c$ of Eq.~(\ref{eq0}).
This test we have done for negative and positive trions,
anticipating that the X$^+$ could show more dependence.
The thin lines were obtained for all hole LLs spaced by the 
same $\Delta\varepsilon=\zeta\hbar\omega_c$.
A more reasonable test is shown with the thicker lines,
obtained for only the (unknown) gaps beyond the $n=1$ LL 
being rescaled by $\zeta$.
The lack of strong dependence justifies the neglect of
exact LL structure for the holes in the following calculation.

\begin{figure}
   \includegraphics[width=3.4in]{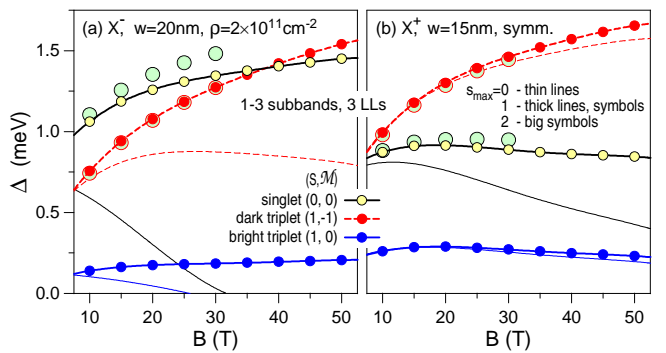}
   \caption{(color online)
      Comparison of magnetic field dependences of negative
      and positive trion binding energies $\Delta(B)$ 
      calculated including five LLs and between one and 
      three subbands.
      (a) X$^-$ in one-sided doped 20~nm GaAs 
      quantum well;
      (b) X$^+$ in symmetric/undoped 15~nm well.}
   \label{fig08}
\end{figure}

Finally, in Fig.~\ref{fig08} we analyze the effect of higher
quantum well subbands.
On two examples (negative trions in a doped 20~nm well and 
positive trions in a symmetric 15~nm well) we compare the
magnetic-field dependence of $\Delta$ calculated using three
LLs and between one and two subbands.
Analogously to the LL mixing, by far the strongest subband 
mixing occurs for the singlet trion.
The effect is especially strong in both systems used here 
as the examples due the strong charge separation in the
lowest subband for the X$^-$ and having two heavy-mass 
particles in the X$^+$.
Evidently, the $s=1$ subband must necessarily be included
in the calculation for all trions in the presence of a doping 
layer, and at least for the singlet state in symmetric or 
undoped wells.
However, the $s=2$ and higher subbands can be neglected
(except only for the singlet trions in wide asymmetric wells 
in high magnetic fields).
Hence, two subbands ($s_{\rm max}=1$) were included in the 
following calculations.

The results of this section can be summarized by two 
conclusions that we believe to be important in the comparison
of numerical and experimental studies of trions.
First, the dark and bright triplet states can be accurately
modeled by exact diagonalization.
The errors due to finite size, limited CI basis, or uncertainty 
in some microscopic parameters either do not affect these trions
or can be eliminated (without need for extraordinary computer
power).
Second, numerical calculations for the singlet trion (by the 
CI exact diagonalization) are far more difficult due to a 
stronger sensitivity to the unknown sample-dependent parameters.
However, note that even for the singlet X$^+$ we reached quite 
satisfactory agreement with the experiment of Vanhoucke {\em 
et al.} (cf.\ Fig.~3 of Ref.~\onlinecite{Vanhoucke01}c).

\subsection{Variational treatment of higher subbands}

As a complementary account of the subband mixing we also used the 
following combination of exact diagonalization (in the plane of 
the well) and the variational approach (in the normal direction).
The pair of model lowest-subband electron and hole wave functions 
\begin{equation}
   \chi(z)=\sqrt{2/w^*}\cos(z\pi/w^*),
\end{equation}
where $w_e^*=w+3.3$~nm and $w_h^*=w+1.75$~nm denote the effective 
layer widths adequate for GaAs/Al$_{0.35}$Ga$_{0.65}$As symmetric 
wells and for $w=10-30$~nm, were each rescaled by an arbitrary 
shrinking factor $\xi\le1$
\begin{equation}
   \chi_\xi(z)=\xi^{-1/2}\chi(\xi^{-1}z).
\label{eqshri}
\end{equation}
The diagonalization was then carried out separately for the $e+h$ 
and $2e+h$ systems and for different combinations of electron and 
hole trial subband states $\chi_\xi$.
The result are the energy maps $E(\xi_e,\xi_h)$, separate for the 
exciton and each trion state.
Having $\xi\ne1$ costs single-particle energy, but stronger 
confinement imposed by $\xi<1$ decreases total Coulomb energy of 
each bound state.
The interpolation and minimization of $E$ with respect to both
variational parameters $\xi$ yield, independently for the X and 
each X$^-$, the ground-state subband parameters 
$(\bar\xi_e,\bar\xi_h)$ and energies $E(\bar\xi_e,\bar\xi_h)$.
The variational trion binding energies were calculated from 
Eq.~(\ref{eqbnd}) and, finally, the correction $\delta\Delta$ 
beyond the lowest-subband approximation was obtained by their 
comparison with the reference values obtained for $\xi_e=\xi_h=1$.

\begin{figure}
   \includegraphics[width=3.4in]{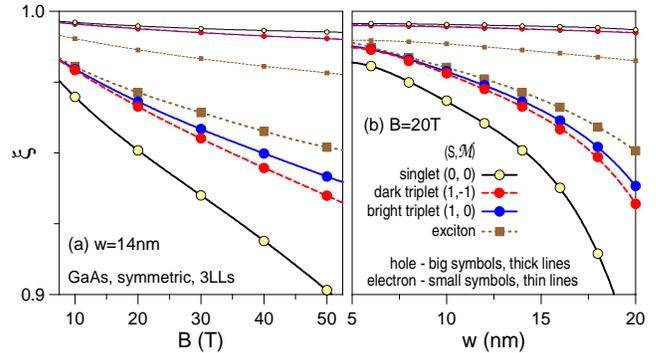}
   \caption{(color online)
      Variational coefficients $\xi$ of Eq.~(\ref{eqshri}),
      rescaling widths of electron and hole wave functions 
      compared to the noninteracting lowest-subband eigenstates,
      calculated for the exciton and trions in an 
      undoped/symmetric GaAs well as a function of magnetic 
      field $B$ (a) and width $w$ (b).}
   \label{fig09}
\end{figure}

In two frames of Fig.~\ref{fig09} we plot the calculated $\xi_e$ 
and $\xi_h$ as a function of $w$ and $B$.
The inaccuracy of the lowest-subband approximation, 
manifested by the interaction-induced distortion (squeeze) 
of the subband wave functions and measured by $\xi$, clearly 
increases as a function of both $B$ and $w$.
Both $\xi_e$ and $\xi_h$ are roughly linear in 
$(w/\lambda)^2\propto w^2B$.
However, $\xi_h$ is considerably smaller than $\xi_e\ge0.98$
(which further enhances the $w_h^*<w_e^*$ asymmetry).
Both $\xi$'s differ between the exciton and different trion 
states.
Remarkably, $\xi_h$ and $\xi_e$ are anti-correlated in a sense 
that $\xi_h$ decreases and $\xi_e$ increases when going from X 
to X$^-_{\rm tb}$, X$^-_{\rm td}$, and X$^-_{\rm s}$ (at any
given $w$ and $B$).

\begin{figure}
   \includegraphics[width=3.4in]{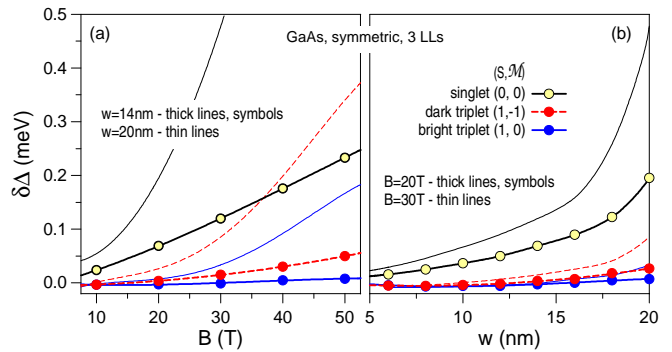}
   \caption{(color online)
      Variational correction $\delta\Delta$ to the trion binding 
      energies due to the squeeze of electron and hole 
      wave functions compared to the noninteracting lowest-subband 
      eigenstates, calculated for undoped/symmetric GaAs 
      wells as a function of magnetic field $B$ (a) and 
      the well width $w$ (b).}
   \label{fig10}
\end{figure}

The squeeze of subband wave functions shown in Fig.~\ref{fig09} 
confirms the main conclusions drawn from inclusion of higher
subbands directly into exact diagonalization (cf.\ Fig.~\ref{fig08}).
It predicts dependence on $w$ and $B$, and explains why the 
subband mixing affects the singlet trion much more than both 
triplets.
The variational correction $\delta\Delta$ beyond the lowest 
subband approximation are plotted in Fig.~\ref{fig10}.
Clearly, the quantitative agreement with exact diagonalization
is rather poor (even if the comparison is made for the same 
$n_{\rm max}=2$ restriction in each method).
The reason is probably the combination of (i) the neglect of 
correlations in the normal direction within the variational 
calculation and (ii) the rather unrealistic choice of the trial 
subband wave function $\chi_\xi$.

\section{Results and discussion}

\subsection{Exciton energy dispersion}

Let us begin the discussion of trion binding energies $\Delta$
with an obvious observation that they strongly depend on the 
energy spectrum of the exciton being captured by a free electron.
This spectrum consists of a continuous in-plane dispersion $E(k)$ 
in each quantized subband.

In an ideal 2D system (lowest LL and zero width), exciton 
dispersion is known exactly,\cite{Bychkov81,Kallin84}
\begin{equation}
   E(k)=-{e^2\over\lambda}\sqrt{\pi\over2}\,
   e^{-\kappa^2}I_0(\kappa^2),
\label{eqKH}
\end{equation}
with $\kappa=k\lambda/2$ and $I_0$ being the modified 
Bessel function of the first kind.
For small $k$, 
\begin{equation}
   E(k)=-{e^2\over\lambda}\sqrt{\pi\over2}\,
   \left(1-\kappa^2+{3\over4}\kappa^4+\dots\right).
\end{equation}
The curvature of $E(k)$ can be attributed to a ``Coulomb 
mass'' $m_{\rm X}(k)=\hbar^2(\partial^2\!E/\partial k^2)^{-1}$,
which must not be confused with the single-particle electron 
or hole cyclotron masses.
For an ideal 2D system, it follows from Eq.~(\ref{eqKH})
\begin{equation}
   m_{\rm X}(k)=\hbar^2
   \left({e^2\lambda\over2}\sqrt{\pi\over2}\right)^{-1}
   \left(1+{9\over2}\kappa^2+\dots\right).
\end{equation}
At zero wave vector, the relation between $m_{\rm X}\equiv 
m_{\rm X}(0)$ and the interaction strength is particularly 
simple,
\begin{equation}
   m_{\rm X}={2\hbar^2\over\lambda^2|E(0)|} 
\label{eqm2d0}
\end{equation}

\begin{figure}
   \includegraphics[width=3.4in]{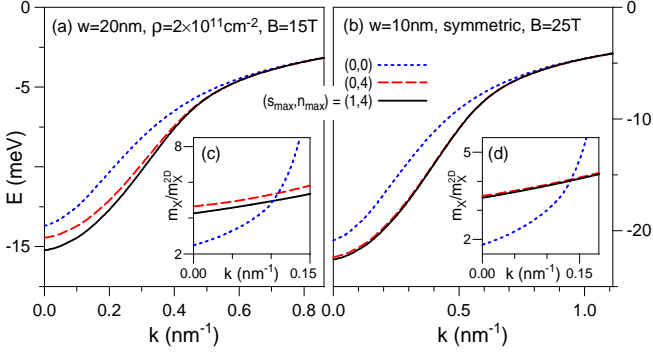}
   \caption{(color online)
      Comparison of excitonic dispersions (energy $E$ vs wave 
      vector $k$) obtained with and without higher LLs and
      subbands, in different GaAs quantum wells.
      Insets: exciton Coulomb mass relative to 
      the ideal value of Eq.~(\ref{eqm2d0}).}
   \label{fig11}
\end{figure}

The magneto-exciton spectra in real quasi-2D structures were 
studied by Lozovik {\em et al.},\cite{Lozovik02} so here we 
concentrate on the finite-width and LL/subband-mixing effects, 
illustrated in Fig.~\ref{fig11} for two different quantum wells 
at different magnetic fields.  
The ground state energy $E(0)$ which defines the maximum 
$e$--$h$ attraction noticeably depends on both LL and 
(in the asymmetric well) subband mixing.
In the insets we plot $m_{\rm X}$ divided by the value given 
by Eq.~(\ref{eqm2d0}) for an ideal 2D system.
In both quantum wells we find significant enhancement of 
$m_{\rm X}$ due to finite well width (by a factor $\sim2$).
Inclusion of higher LLs causes additional increase by another 
factor of $\sim2$, while the subband mixing has a much 
weaker effect.

\subsection{Trion binding energies}

\begin{figure}
   \includegraphics[width=3.4in]{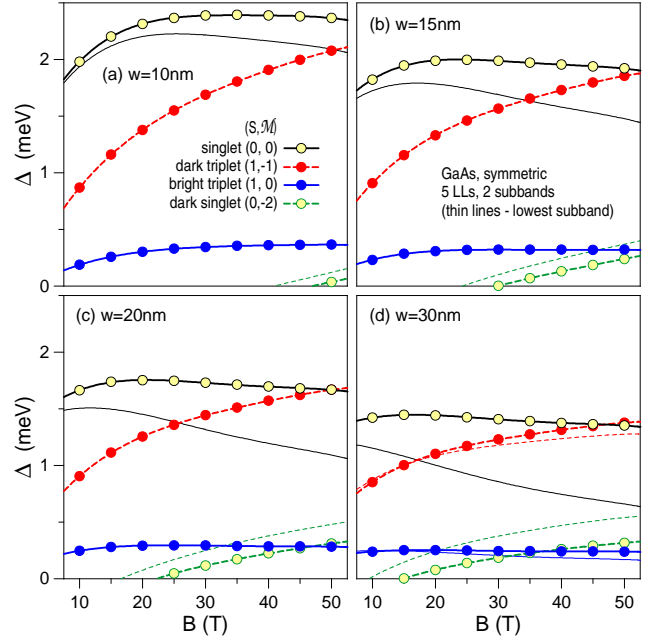}
   \caption{(color online)
      Dependence of the trion binding energies $\Delta$ 
      on magnetic field $B$, for undoped or symmetric
      GaAs quantum wells of different widths $w$.
      Different bound trion states are distinguished by 
      the $2e$ spin $S$ and relative angular momentum 
      $\mathcal{M}$. 
      Five LLs and up to two subbands were included in exact 
      diagonalization on a sphere.}
   \label{fig12}
\end{figure}

Let us now turn to the presentation of our main numerical results,
that is to the dependence of the trion binding energies $\Delta$ 
in one-sided doped GaAs/Al$_{0.35}$Ga$_{0.65}$As quantum wells on
$w$, $\varrho$, and $B$.
The $\varrho=0$ data are also adequate for the symmetrically doped 
wells.
The calculations were carried out for $2Q=20$ and including two 
subbands ($s_{\rm max}=1$) and five LLs ($n_{\rm max}=4$).
The accuracy of this basis was discussed in Sec.~\ref{secCaA}.

Plotted values of $\Delta$ are the Coulomb binding energies.
The appropriate Zeeman terms must be additionally included to 
determine the absolute trion ground state or the energy splittings 
in the photoluminescence spectra.

Depending on the parameters, we found up to four different trions 
with significant binding energy in the energy spectra similar to 
Fig.~1 of Ref.~\onlinecite{Wojs00}.
Besides the ``singlet'' X$^-_{\rm s}$ with $(S,\mathcal{M})=(0,0)$, 
``dark triplet'' X$^-_{\rm td}$ (1,-1), ``bright triplet'' 
X$^-_{\rm tb}$ (1,0), an additional ``dark singlet'' X$^-_{\rm sd}$ 
(0,-2) was identified in some systems.
Their binding energies will be denoted by $\Delta_{\rm s}$,
$\Delta_{\rm td}$, $\Delta_{\rm tb}$, and $\Delta_{\rm sd}$.

In the following figures, the bright and dark states are 
distinguished by solid and dashed curves.
On the other hand, the singlets and triplets are marked by
open and full dots, respectively.
The sets of curves obtained in the lowest-subband approximation 
(LSA) are also drawn for comparison with thin lines without
the dots.

\subsubsection{Dependence on magnetic field}

\begin{figure}
   \includegraphics[width=3.4in]{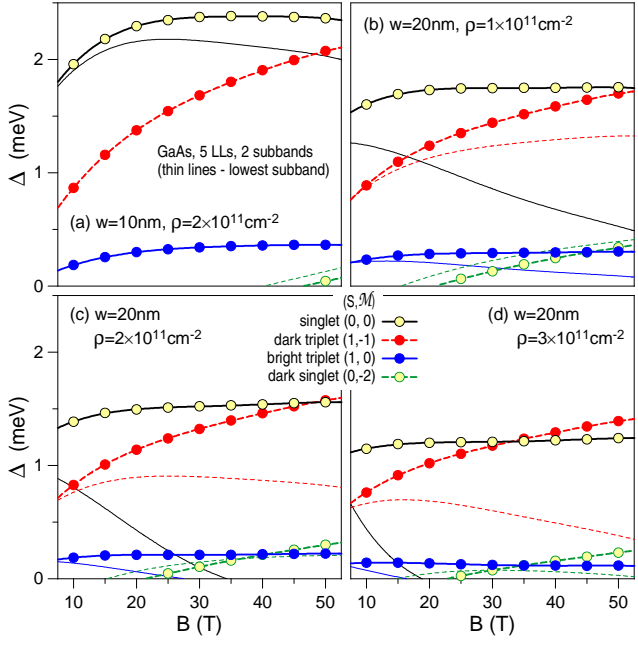}
   \caption{(color online)
      Similar to Fig.~\ref{fig12}, but for one-sided doped 
      $w=10$ and 20~nm GaAs quantum wells of different  
      electron concentration $\varrho$.}
   \label{fig13}
\end{figure}

We begin with the magnetic-field dependence shown in 
Fig.~\ref{fig12} for four symmetric quantum wells of different
width and in Fig.~\ref{fig13} for two doped wells of different
width and electron concentration.
The singlet remains the most strongly bound state in all frames, 
with the singlet-triplet crossing pushed to much higher magnetic 
fields than anticipated by the earlier LSA calculations.
Especially in the wide ($w=20$~nm) doped wells the LSA fails 
completely, with the actual $\Delta_{\rm s}(B)$ far more 
resembling that in the symmetric well.
In all systems, $\Delta_{\rm s}(B)$ and $\Delta_{\rm tb}(B)$ 
rise at weak magnetic fields, but the increase saturates at
$B=15-30$~T (depending on $w$ and $\varrho$), beyond which 
both $\Delta_{\rm s}(B)$ and $\Delta_{\rm tb}(B)$ remain 
essentially flat at $\Delta_{\rm s}=1.5-2.4$~meV and 
$\Delta_{\rm tb}=0.2-0.4$~meV depending on $w$ and $\varrho$.
In contrast, $\Delta_{\rm td}(B)$ largely retains the 
characteristic $\sqrt{B}$ dependence of an ideal 2D
system ($w=0$ and lowest LL).
The weakly bound dark singlet emerges in wider wells 
and at higher fields, with a monotonically increasing 
$\Delta_{\rm sd}(B)$.

\subsubsection{Dependence on electron concentration}

\begin{figure}
   \includegraphics[width=3.4in]{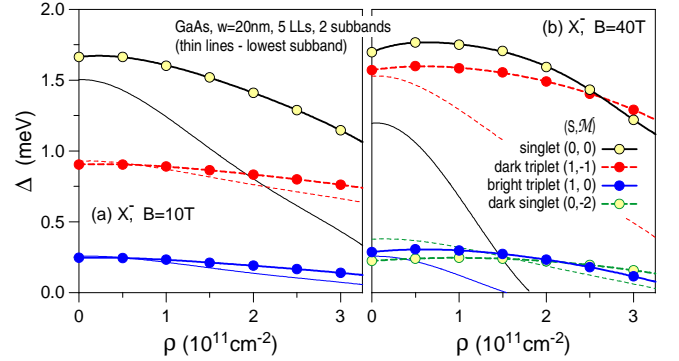}
   \caption{(color online)
      Similar to Fig.~\ref{fig12}, but with trion binding 
      energies $\Delta$ plotted as a function of electron 
      concentration $\varrho$, for a 20~nm GaAs well at 
      fields $B=10$~T (a) and 40~T (b).}
   \label{fig14}
\end{figure}

In Fig.~\ref{fig14} we study in more detail the dependence 
on electron concentration $\varrho$.
It was clear from Fig.~\ref{fig12} that the effect is 
negligible for narrow wells, so we only plot the data for
$w=20$~nm (with a relatively low $B=10$~T and a very high
$B=40$~T compared in the two frames).
Of all states, only the singlet shows a noticeable dependence 
on $\varrho$.
However, it must be realized that if $\Delta$ were to be 
understood as the binding energy of a well defined bound 
state, at most weakly perturbed by the surrounding electrons,
then $B$ and $\varrho$ must be restricted to the values 
describing a sufficiently ``dilute'' system.
A convenient measure is the LL filling factor $\nu=2\pi
\varrho\lambda^2$ (proportional e.g.\ to the area of the 
dark triplet trion divided by the average area per electron).
While for $B=40$~T the system can be regarded dilute 
$\nu<1/3$ throughout the frame, at $B=10$~T the 
lowest LL fills completely ($\nu=1$) already at $\varrho
=2.4\cdot10^{11}$~cm$^{-2}$.

It is clear from Fig.~\ref{fig14} that the LSA fails completely 
to describe the effect of charge separation in asymmetric wells.
It appears that, except for the extreme cases of highly doped 
wide wells, the effect can be simply ignored. 
However, if it need be included, the calculation must take 
into account the mixing with at least one higher subband.

\subsubsection{Dependence on quantum well width}

\begin{figure}
   \includegraphics[width=3.4in]{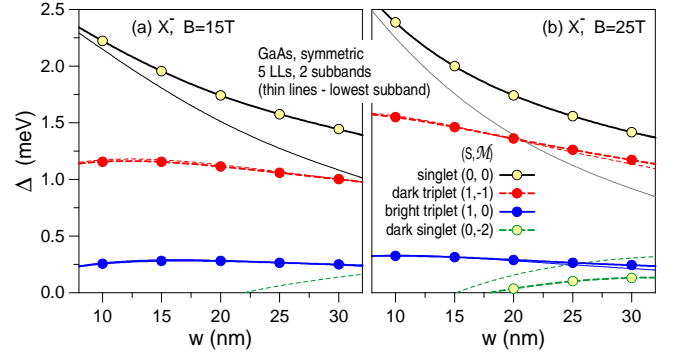}
   \caption{(color online)
      Similar to Fig.~\ref{fig12}, but with trion binding 
      energies $\Delta$ plotted as a function of width $w$ 
      of a symmetric GaAs quantum well, at fields $B=15$~T 
      (a) and 25~T (b).}
   \label{fig15}
\end{figure}

In Fig.~\ref{fig15} we plot $\Delta$ as a function of the 
width $w$ (of a symmetric well) at two different magnetic
fields.
The LSA works remarkably well for the two triplet states,
neither of which shows a significant $\Delta(w)$ dependence 
throughout the wide range of $w=10-30$~nm.
In contrast, and in good agreement with the experiment of 
Vanhoucke {\em et al.} (cf.\ Fig.~5 of 
Ref.~\onlinecite{Vanhoucke01}a), $\Delta_{\rm s}(w)$ 
decreases markedly in the same range, although nearly not 
as much as predicted by the LSA.

\begin{figure}
   \includegraphics[width=3.4in]{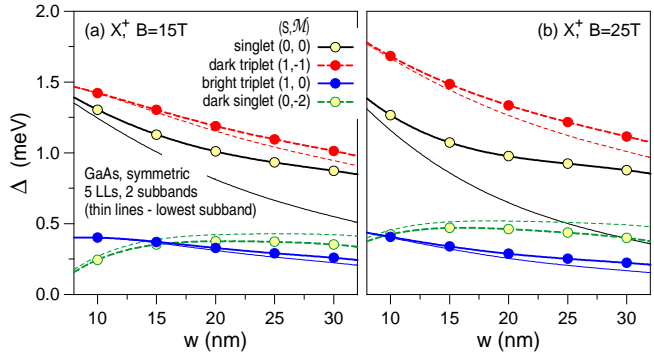}
   \caption{(color online)
      Same as Fig.~\ref{fig15}, but for positive trions.}
   \label{fig16}
\end{figure}

In Fig.~\ref{fig16} we use the width dependence $\Delta(w)$ to
illustrate the difference between negative and positive trions.
Remarkably, at $B\ge15$~T dark triplet is the most strongly 
bound X$^+$ throughout the entire shown width range $w=10-30$~nm.
Only the singlet positive trion has a lower binding energy
than its negative counterpart.
All other trions enhance their binding when going from X$^-$ 
to X$^+$.
This is especially true of the dark singlet which emerges
as an additional robust bound state in the X$^+$ spectrum.
Also, the width dependence for the pair of triplets is more 
pronounced in the X$^+$.
Similarly as in X$^-$, the LSA works well for all positive 
trions except for the bright singlet.

\subsubsection{Singlet-triplet crossing}

From the comparison of binding energies of different trions 
one can try to establish the absolute trion ground state in 
a given quantum well ($w$ and $\varrho$) at a given field $B$.
This might be especially important in the context of 
``quasiexcitons'',\cite{Wojs06} proposed to form in incompressible
electron fluids, such as the Laughlin fractional quantum Hall 
state at LL filling factor $\nu=1/3$.
Only those quasiexcitons formed from a dark triplet trion 
were found to cause discontinuity of the PL emission energy 
at $\nu=1/3$, suggesting that an appropriate structure design 
(to achieve significant occupation of the dark triplet state) 
might be essential for optical detection of the Laughlin 
incompressibility.

\begin{figure}
   \includegraphics[width=3.4in]{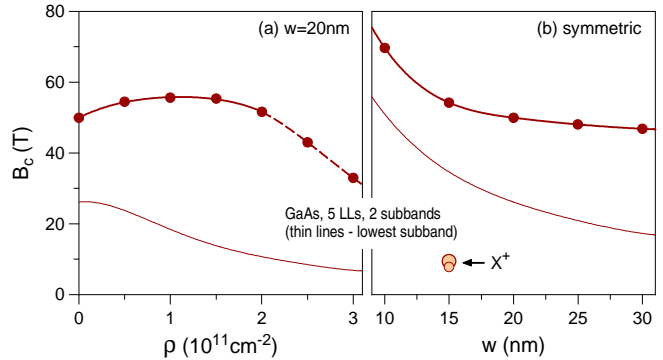}
   \caption{(color online)
      Critical magnetic field $B_c$ for the crossing 
      of Coulomb binding energies of the
      singlet and dark-triplet states of the negative 
      trion, as a function of electron concentration 
      $n$ (a) and quantum well width $w$ (b).
      Open circles in (b) mark $B_c$ for a positive
      trion in a $w=15$~nm well.}
   \label{fig17}
\end{figure}

In Fig.~\ref{fig17} we plot the critical magnetic field 
$B_c$ at which the singlet-triplet crossing occurs in 
the trion Coulomb spectrum, as a function of both $w$ 
and $\varrho$.
An accurate estimate of $B_c$ is quite difficult due to 
accumulation of errors in the two compared $\Delta$'s.
Nonetheless, it is clear that inclusion of subband mixing
favors the singlet trion, pushing $B_c$ to much higher 
values.

The fact that the LSA underestimates $B_c$ was found by 
Whittaker and Shields\cite{Whittaker97} (for symmetric wells).
It can be explained by the same perturbative argument we 
invoked earlier to explain different effect of the LL mixing 
on the singlet and triplet states: by noting that Coulomb 
interactions are stronger in the singlet state, causing 
more efficient mixing of excited single-particle states.
However, in contrast to Whittaker and Shields, and in 
agreement with experiment of Vanhoucke {\em et al.} 
(cf.\ Figs.~5 and 6 of Ref.~\onlinecite{Vanhoucke01}b),
we find that $B_c$ {\em decreases} as a function of $w$ 
(e.g., leading to a crossing below 50~T in a 30~nm well).

Other conclusions from Fig.~\ref{fig17} are the following.
Doping of even a fairly wide, 20~nm well has small effect 
on $B_c$ up to $\varrho\sim2\cdot10^{11}$~cm$^{-2}$ (also, 
its decrease at higher concentrations is most certainly an 
artifact of the $s\le1$ approximation; cf.\ Fig.~\ref{fig08}).
The crossing of Coulomb binding energies calculated here
is different from the crossing of the ground state energies 
due to additional electron Zeeman energy that must be added
for the singlet (to lower $B_c$ compared to our plot).
It is also different from the crossing of recombination 
energies in the PL spectra, due to dependence of the hole 
Zeeman energy on a particular exciton or trion state.

In Fig.~\ref{fig17}(b) we also marked $B_c$ for a positive 
trion in a symmetric 15~nm well (small circle for LSA;
large circle for $s\le1$).
In good agreement with the experimental reports, it is much 
smaller than the value for X$^-$, predominantly due to the 
opposite effective-mass ratio.

\subsubsection{Effect of nearby ionized impurities}

Being charged, trions are efficiently localized by even 
fairly remote ionized impurities.
Especially for the dark triplet, this might appear as a 
possible mechanism for breaking translational invariance 
of an otherwise perfect quantum well, needed for the 
explanation of experimentally observed optical recombination 
of this state.
However, Dzyubenko {\em et al.}\cite{Dzyubenko06} showed 
recently that this mechanism is not efficient.
The ${\rm D}^+{\rm X}^-_{\rm td}$ state was found stable 
only at sufficiently large distances $d$ of the ionized 
donor D$^+$ from the center of the quantum well, at which 
its oscillator strength remains small ($\sim1$\% of the 
X$^-_{\rm s}$).

\begin{figure}
   \includegraphics[width=3.4in]{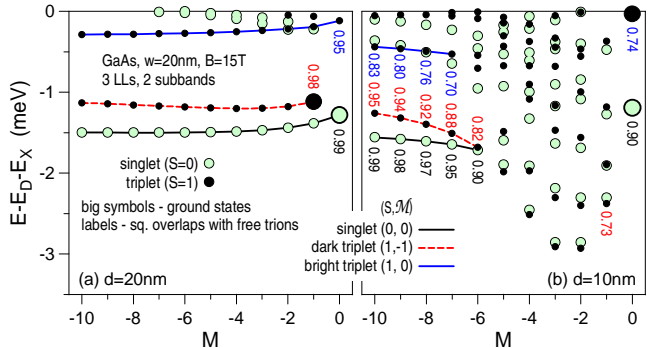}
   \caption{(color online)
      Energy spectra of the $2e+h$ system in the presence of 
      an ionized donor D$^+$ at a distance $d=w$ (a) and 
      $d=w/2$ (b) from the well center (energy $E$, 
      measured relative to an exciton unbound from the D$^0$, 
      plotted as a function of angular momentum on a plane $M$), 
      calculated for a symmetric/undoped $w=20$~nm GaAs 
      quantum well at magnetic field $B=15$~T, including 
      three LLs and two subbands.
      Lines connect the D$^+{\rm X}^-$ states formed from 
      different trions; the $S=0$ and 1 ground states are
      marked with bigger symbols; and the color labels 
      indicate squared overlaps with the corresponding 
      free-trion states $|\left<\right.\!{\rm X}^-|
      {\rm D}^+{\rm X}^-\left.\!\right>|^2$.}
   \label{fig18}
\end{figure}

We repeated this calculation in our model, but chose a wider, 
$w=20$~nm quantum well and a higher field $B=15$~T, 
in anticipation of a stronger effect of subband mixing induced 
by an off-center ($d\ne0$) point charge.
In Fig.~\ref{fig18} we compare the $2e+h$ energy spectra in 
the presence of a D$^+$ at $d=w$ and $w/2$.
On the horizontal axis is the total angular momentum converted 
to the plane, $M=L_z-Q$ (neither $L$ nor $\mathcal{M}$ are 
conserved in the presence of D$^+$, leaving $L_z$ or $M$ as 
the only good orbital quantum number).
The energy $E(M)$ is measured relative to the configuration
consisting of a ground state exciton unbound from a neutral 
donor state D$^0={\rm D}^++e$.
The plotted energy difference is hence the negative of
\begin{equation}
  \Delta(M)=E_{\rm X}+E_{\rm D}(M)-E(M),
\end{equation}
which is the trion binding energy in the presence of a D$^+$,
generalized from Eq.~(\ref{eqbnd}).
The singlet and triplet ground states (states with the lowest 
$E$) are also indicated.

The average trion-donor distance depends on both $d$ and $M$.
At sufficiently large distances of the donor from the well 
(in our example, at $d\ge w$), the trion is captured without 
a significant distortion of its wave function.
The lowest state for each trion occurs at $M=\mathcal{M}$ 
(trion's relative angular momentum).
Each trion's $\Delta$ is slightly decreased, but the 
dependence on $d$ or $M$ is weak in this regime.
In contrast, when the donor approaches the well, new trions 
bind and mix with X$^-_{\rm s}$, X$^-_{\rm td}$, and 
X$^-_{\rm tb}$.
The effect disappears at large $|M|$, but it is also 
relatively weak at $M=0$, because the additional trions 
have $\mathcal{M}\ne0$.\cite{Dzyubenko06}
As pointed out in Ref.~\onlinecite{Dzyubenko06}, for $d=w/2$ 
(donor at the wall of the quantum well), $\Delta=1.2$~meV of 
the singlet ground state is only slightly smaller than 
$\Delta=1.55$~meV of a free X$^-_{\rm s}$, but the triplet 
ground state essentially unbinds.

\section{Conclusion}

We used exact numerical diagonalization on a sphere to 
calculate the energy spectra of negative and positive 
trions in doped GaAs quantum wells in high magnetic fields.
The results obtained with the inclusion of up to three
quantum well subbands were compared with a different method,
combining exact diagonalization of the in-plane dynamics 
with a variational calculation for the normal direction.
The obtained dependences of the trion binding energies 
$\Delta$ on the quantum well width $w$, electron 
concentration $\varrho$, and magnetic field $B$ 
significantly improve over previous theoretical estimates.
In particular, the symmetric-well and lowest-subband 
approximations are both invalidated for a wide class 
of realistic systems.
Presented detailed analysis of the accuracy and convergence 
establishes the exact diagonalization on a sphere for future 
quantitative studies of excitonic complexes.
For example, it can be used for the analysis of exciton 
and trion wave functions ($e$--$e$ and $e$--$h$ correlations, 
oscillator strengths, LL/subband projections, etc.) or
more detailed studies of the interaction of excitonic 
complexes with ionized impurities and/or free carriers.

\section*{Acknowledgment}

Work supported by grants DE-FG 02-97ER45657 of US DOE and 
N20210431/0771 of the Polish MNiSW.

\end{document}